\documentclass[10pt,a4paper,reprint,aps]{revtex4-2}
\usepackage[utf8]{inputenc}
\usepackage{amsmath}
\usepackage{amsfonts}
\usepackage[margin=0.5in]{geometry}
\usepackage{amssymb}
\usepackage{graphicx}
\usepackage{xcolor}
\usepackage{caption}
\usepackage{subcaption}
\usepackage{color, colortbl}
\definecolor{Gray}{gray}{0.8}

\begin{document}

\title{\textbf{Theoretical analysis of cargo transport by catch bonded
motors in optical trapping assays}}






\author{Naren Sundararajan$^{1,2,*}$, Sougata Guha$^{2,*}$, Sudipto Muhuri$^{1,\dagger}$, Mithun K. Mitra$^{2,\ddagger}$\\
\small{\emph{$^1$Department of Physics, Savitribai Phule Pune University, Pune, India\\
$^2$Department of Physics, IIT Bombay, Mumbai, India\\
$^*$These authors contributed equally\\
Email : $^\dagger$sudipto@physics.unipune.ac.in, $^\ddagger$mithun@phy.iitb.ac.in
}}}

\begin{abstract}

Dynein motors exhibit catch bonding, where the unbinding rate of the motors from microtubule filaments decreases with increasing opposing load. The implications of this catch bond on the transport properties of dynein-driven cargo are yet to be fully understood. In this context, optical trapping assays constitute an important means of accurately measuring the forces generated by molecular motor proteins.
We investigate, using theory and stochastic simulations, the transport properties of cargo transported by {\em catch bonded} dynein molecular motors - both singly and in teams - in a harmonic potential, which mimics the variable force experienced by cargo in an optical trap. 
We estimate the biologically relevant measures of {\it first passage time} - the time during which the cargo remains bound to the microtubule and {\it detachment force} -the force at which the cargo unbinds from the microtubule, using both two-dimensional and one-dimensional force balance frameworks. Our results suggest that even for cargo transported by a single motor, catch bonding may play a role depending on the force scale which marks the onset of the catch bond. By comparing with experimental measurements on single dynein-driven transport, we estimate realistic bounds of this catch bond force scale. Generically, catch bonding results in increased persistent motion, and can also generate non-monotonic behaviour of first passage times. For cargo transported by multiple motors, emergent collective effects due to catch bonding can result in non-trivial re-entrant phenomena wherein average first passage times and detachment forces exhibit non-monotonic behaviour as a function of the stall force and the motor velocity.

\end{abstract}

\maketitle


\section{Introduction}
The process of intracellular transport involves the collective action of many motor proteins \cite{alberts1998cell,welte1998cell,howard2002mechanics,welte2004currbio,ferenz2009currbio,muhuri2010pre,driver2010jccp,bhat2012physbio,mckinley2012jtheobio,sutradhar2014jtheobio,chandel2015epl,mclaughlin2016softmatter,puri2019prr,campas2008biophysj,brugues2009prl}. In particular kinesins and dyneins walk along microtubule filaments to transport cargo. While kinesin-driven motion has been widely studied and characterised, both in experiments and theory, a crucial difference between kinesin and dynein lies in their unbinding characteristics from the MT filament. Kinesins are known to exhibit slip bonding \cite{pyrpassopoulos2017molbiocell,uemura2002pnas,andreasson2015elife,coppin1997pnas}, although recent experiments show that the unbinding characteristics for kinesin are more nuanced than previously understood \cite{khataee2019prl}. For slip bonds, the unbinding rates increase with increasing load forces. In contrast, strikingly, dyneins exhibit catch bonding, where in certain force regimes, the unbinding rate of the dynein motor decreases with increasing opposing load \cite{kunwar2011pnas,leidel2012biophysj,rai2013cell}. This counter-intuitive unbinding characteristic can have profound implications on the transport properties of dynein-driven cargo \cite{chandel2015epl,puri2019prr}.



Optical trap assays provide an experimental means to measure and quantify force generated by individual motors at the scale of  single-molecule resolution \cite{leidel2012biophysj, rai2013cell}. Since, in most optical trap setups, the restoring force experienced by a motor increases with the distance of the cargo from the trap center, it should be modelled as a variable force ensemble, with the motion being arrested when the cargo has travelled a sufficient distance such that the force on the motors exceeds the stall force. In contrast, most theoretical and simulation studies have investigated the transport characteristics of cargo in the presence of a constant load force \cite{klumpp2005pnas, nair2016pre,bhat2016epje}. Experiments in optical traps have enabled systematic calibration of the motility characteristics of the transported cellular cargo on variation of individual motor properties \cite{rai2013cell, mallik2013ticb}. These experiments have also served to highlight how the different collective transport characteristics of cellular cargo that are carried by kinesin as compared to dynein motors. In particular, it has been observed that the tenacity of a team of motors, defined as the time spent above the half maximal force, increases linearly with motor number for dynein, while it is independent of motor number for kinesin \cite{rai2013cell}. It has been argued that in fact the {\it catch bond mechanism} at play for dynein bond under load force leads to increased bond lifetime, which in turn allows the trailing motors to catch up and share the opposing load force \cite{rai2013cell, mallik2013ticb}.

\begin{figure*}[t]
    \centering
    \includegraphics[width=\linewidth]{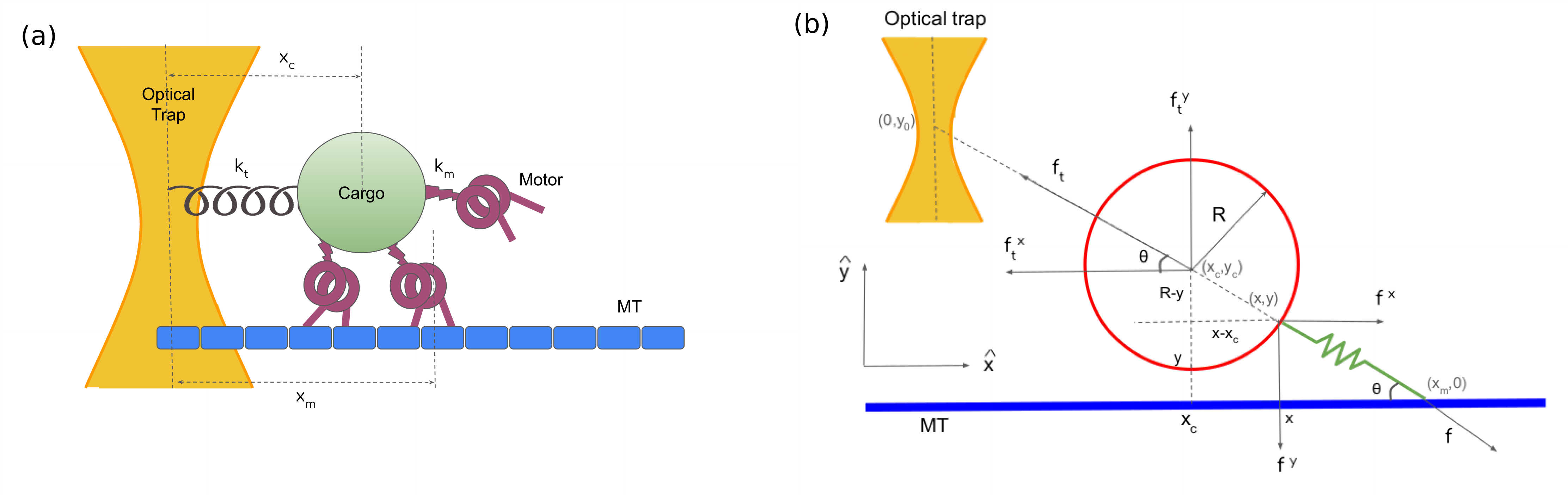}
    \caption{Schematic of multiple motor-driven cargo transport on a microtubule (MT) filament in an optical trap. (a) The optical trap and the motors are denoted as harmonic springs with spring constants $k_t$ and $k_m$ respectively. The positions of the cargo and the motors from the optical trap centre are denoted by $x_c$ and $x_m$ respectively. (b) 2D schematic of the force-balanced situation for a motor transport under optical trap. The bead of radius $R$ is moving in x-y plane and the MT is along x-direction. The optical trap center is at $(0,y)$ while the cargo position is denoted by ($x_c,y_c$). The motor is attached to the bead at ($x,y$) while the other end of the motor is attached to the MT at ($x_m,0$) and makes an angle $\theta$ with the MT. The cargo is experiencing two forces - force due to motor extension ($f$) and trap force ($f_t$). The components of these two forces along x and y-directions are denoted as $f^x$, $f^y$, $f_t^x$ and $f_t^y$ respectively.}
    \label{fig:schematic}
\end{figure*}

While it was thought that the primary functional role of the biological catch bond was restricted to improving surface adhesion properties of bacteria and cells when subjected to external forces or flow field \cite{piel2009currbio}, many recent  theoretical studies have indicated  a much wider array of functional role of the biological catch bonding in intracellular processes ranging from regulating intracellular transport \cite{hancock2014natrevmolcellbio,nair2016pre,puri2019prr}  to generating oscillations in muscle fibres and mitotic spindles during cell division \cite{guha2021biophysj}. Thus it is imperative to understand the underlying mechanism which governs the collective behaviour of catch bonded dynein motors.
In this context, experiments using optical trap assay provide a useful means of not only investigating the role of catch bond in dynein motors on intracellular properties but it may also be used for systematically studying the effects of variation of the single motor properties such as the stall force and the motor velocity which are known to regulate the overall motility characteristics of the transported cellular cargo \cite{ohashi2019traffic,belyy2016natcellbio,ross2006natcellbio,torisawa2014natcellbio,hirakawa2000pnas,hong2016biophysj,schlager2014cellrep,schlager2014emboj,chaudhary2018traffic}. 


In this paper, we seek to understand the role of catch bond in mediating transport of cellular cargo. To this end, we study the transport properties of cargo carried both by single dynein as well as by multiple dynein motors in a variable force optical trap. 



\section{Model and Methods}
We model the transport of a cellular cargo that is subjected to force due to motors that are attached to cargo and walk on microtubule filaments, as well as an opposing restoring force arising due to the harmonic potential of the optical trap. The motors themselves stochastically (un)bind (from)to the MT filament in a force-dependent manner. For non-catch bonded motors, the unbinding rate increases exponentially with the load force - a characteristic of {\it slip bond} \cite{bell1978models,zhurkov1965kinetic}, such that 

\begin{equation}
    \epsilon (f) = \epsilon_{o} e ^{f/f_k}
    \label{eq:k-slip}
\end{equation}
where $f_k$ is the characteristic detachment force and $\epsilon_o$ is the load-free unbinding rate for slip-bonded motor.

In contrast, for catch bonded dynein motors, it has been observed that beyond a certain threshold load force, the unbinding rate is a decreasing function of the force - a signature of catch bonding behaviour exhibited by dynein motors \cite{kunwar2011pnas}. We denote the corresponding threshold force at which catch bonding is activated in dynein as $f_m$. It is worthwhile to point out that $f_m$ need not necessarily be same as the stall force $f_s$ of the motors. The unbinding rate for single dynein motor can be modelled and fitted using a phenomenological TFBD model \cite{nair2016pre} which reproduces the observed experimental unbinding rates of cytoplasmic dynein motor. Within the framework of the TFBD model, the unbinding rate of single dynein motor can be expressed as,


\begin{equation}
    \epsilon (f) = \epsilon_{o} \exp [- E_d(f) + f/f_d ~]
    \label{eq:catch-ku}
  \end{equation}
where $f_d$ is the characteristic detachment force of dynein motor  in the slip region ($f < f_m$). The deformation energy $E_d$ represents the influence of catch bond behaviour, which sets in  when $ f > f_m$, and is expressed as ~\cite{nair2016pre},
\begin{equation}
    E_d(f) = \Theta (f  - f_m)~ \alpha \left[1 - \exp\left(-\frac{f - f_m}{f_o}\right)\right]
    \label{eq:deform}
\end{equation}
where the $\Theta(x)$ denotes the Heaviside function which is equal to 1 for $x>0$ and zero otherwise. The parameter $\alpha$ is a measure of catch bond strength and is measured in units of $k_BT$. The parameter $f_o$ is associated with the force scale corresponding to the deformation energy due to catch bond. 


\begin{table*}
\label{tab:Table}
\begin{center}
\begin{tabular}{ |c|c|c| }
\hline
\rowcolor{Gray}
Parameter                           & Symbol       & Value (Unless otherwise mentioned) \\
\hline
Binding rate                        & $\pi_o$        & 1.6 $s^{-1}$  \cite{king2000dynactin,muller2008tug} \\ \hline
Unbinding rate (under no load)        & $\epsilon_o$ & 0.27 $s^{-1}$  \cite{mallik2005currbio,muller2008tug} \\ \hline
Principal velocity ( under no load) & $v_o$        & 362 $nm$ $s^{-1}$ \cite{mallik2005currbio} \\ \hline
Motor spring restlength             & $l_o$        & 55 $nm$  \cite{mckenney2010cell} \\ \hline
Stall force for Dynein              & $f_s$        & 1.1 $pN$  \cite{rai2013cell,muller2008tug} \\ \hline
Detachment force for Dynein         & $f_d$        & 0.67 $pN$  \cite{nair2016pre} \\ \hline
Catch bond threshold force          & $f_m$        & chosen same as $f_s$  \cite{puri2019prr} \\ \hline
Deformation force scale             & $f_o$        & 7 $pN$  \cite{puri2019prr} \\ \hline
Catch bond Strength / Deformation energy scale             & $\alpha$        & 0 - 40 $k_B T$          \cite{puri2019prr} \\ \hline
Motor spring stiffness              & $k_m$        & 0.1 - 0.3 $pN$ $nm^{-1}$   \cite{sakakibara1999nature} \\ \hline
Trap Stiffness                      & $k_t$        & 0.01 -0.03 $pN$ $nm^{-1}$  \cite{rai2013cell,brenner2020sciadv} \\ \hline
\end{tabular}
\end{center}
\caption{The table contains the symbols of physical parameters and their values that are used throughout the manuscript (unless mentioned otherwise).}
\end{table*}

For cargo transported by a single motor, we can calculate the runtime and detachment force distributions analytically. We define $S(t)$ as the {\it Survival probability} distribution of the cellular cargo i.e., the probability that the cargo remains attached to the filament (through the bound motor) after time $t$, starting from the initial position of the optical trap center at time $t=0$. Then for a small time interval $\Delta t$, $S( t + \Delta t) = S(t)[ 1 - \epsilon(f) \Delta t ]$ and corresponding time evolution of $S(t)$ is,

\begin{equation}
    \frac{dS}{dt} = -\epsilon(f)S
    \label{eq:dS(t)}
\end{equation}
which yields a solution of $S(t)$, which has the general form,

\begin{equation}
    S(t) = \exp \left [ -\int_{0}^{t} \epsilon dt^{'}\right ]
    \label{eq:S(t)}
\end{equation}

The probability distribution of the survival time, $S(t)$, can then be obtained by substituting the expression for $\epsilon$ in the integral form of Eq.(\ref{eq:S(t)}). An experimentally accessible measure is the first passage time defined as the total time the cargo remains bound to the MT before motor that tethers it to the filament, unbinds. The {\it First passage time} distribution, $F(t)$, can then easily be calculated from Eq.(\ref{eq:dS(t)}) using the relation $F(t)=-\frac{dS}{dt}$.  Another quantity of experimental interest is the detachment force which is defined as the force exerted on the cargo at the time of unbinding of the motor from the MT filament. Once we obtain the first passage time distribution, we can calculate the detachment force distribution, $P(f)$, using the following expression -
\begin{equation}
    P(f)=\frac{\int_0^{t} f(t')F(t') dt'}{\int_0^{t} F(t')dt'}
\label{eq:P_f}
\end{equation}

We also perform stochastic simulations for the cargo trajectories both for the case of transport of cargo by a single motor and by multiple motors. (See Supplementary Section 1 for simulation method details).
\\

\section{Results}
\subsection{Cargo transport by a single motor}
In order to investigate the effect of catch bonded behaviour on the transport characteristics, we first focus on the case of a cargo being transported by a single dynein motor in an optical trap. To characterise the motion, we focus on  experimentally accessible measures of  first passage time (FPT) and detachment force distribution of the cargo. The cargo position at any instant of time is determined by the force balance condition on the cargo.  

\subsubsection{Force balance in two-dimensions}
For transport of a cellular cargo by a single motor, the cargo is subjected to pulling force of a motor that is attached to the underlying microtubule filament and an opposing restoring force arising from the potential due to optical trap. In general, the motion of the cargo as it moves away from the optical trap center, has components both along the horizontal direction (along the axis of microtubule length) as well as the vertical direction. While many previous studies on the motor transport in optical trap setting have only considered one-dimensional transport of the cargo \cite{kunwar2011pnas, rai2013cell, klumpp2005pnas}, ignoring the vertical component of the forces due to the motor and the optical trap, in reality, the cargo is subjected to force along the vertical direction as well \cite{svoboda1994cell,pyrpassopoulos2020biophysj,khataee2019prl} (see Fig. \ref{fig:schematic}b).
The effect of the optical trap on the cellular cargo is modelled as a harmonic potential. While the ratio of trap stiffness in the horizontal and vertical directions varies in range of $1.5-4$ depending on the choice of the bead size and its coating material \cite{ashkin1992biophysj,bormuth2008opticsexp}, for simplicity, we assume that the harmonic potential is isotropic with a spring constant $k_t$. We consider a spherical cargo of radius $R$. We choose the coordinate system such that the MT is positioned along $y=0$ and the trap center is set at position $(0,y_o)$. We assume that the cargo starts from the trap center while the motor attaches to the MT at (0,0) at time $t=0$. Note that molecular motors typically have a finite rest length $l_o$, defined as the maximum extension of the motor for which the motor experiences no load. In the initial configuration, we assume $y_o=R+l_o$ such that the motor is in a vertical position with an extension equal to $l_o$. As the motor starts walking, it stretches beyond the rest length and starts experiencing a force from the optical trap. At any time instant, the motor binding position on the cargo as $(x,y)$, and the cargo centre is denoted as $(x_c,y_c)$.

Force balance condition for the cargo along the horizontal and vertical direction implies (Fig. \ref{fig:schematic}b),

\begin{eqnarray}
  k_t x_c &=& k_m\Theta(l_m-l_o) \left[ (x_m-x) - l_o\frac{(x-x_c)}{R}\right] \label{eq:xc} \\
    k_t(y_o-y_c) &=& k_m\Theta(l_m-l_o) \left[y-l_o\frac{(y_c-y)}{R}\right] \label{eq:yc}
\end{eqnarray}

where $l_m$ is the length of the motor. It is worthwhile to point out that the above force balance condition between motor force and trap force in vertical direction (Eq. \ref{eq:yc}) is valid as long as the cargo does not touch the underlying MT. Once the cargo touches the MT, an additional normal force (along the vertical direction) acts on the cargo, which must also be accounted for in the force balance condition \cite{fisher2005pnas,khataee2019prl,pyrpassopoulos2020biophysj}.

Additionally, from geometrical considerations, the motor binding position $(x,y)$ must satisfy 
\begin{equation}
    (x-x_c)^2+(y-y_c)^2 = R^2 \label{eq:surface}
\end{equation}

Finally for equilibrium,  we must have,
\begin{equation}
    tan\theta = \frac{y}{x_m-x} = \frac{y_c}{x_m-x_c} \label{eq:triangle}
\end{equation}



With these force-balance conditions and geometric constraints, we perform stochastic simulations of a single dynein-driven cargo transport under optical trap (see Suppl. Sec. 1 for details). The motor position $(x,y)$ and cargo position $(x_c,y_c)$ allow us to compute the motor extension and hence the force $f$ experienced by the motor. The motor can then either detach from the MT with the unbinding rate $\epsilon (f)$ (Eq. \ref{eq:catch-ku}) or move along the MT with a velocity $v_m(f)$. For simplicity, we assume a linear force-velocity relation for the motor \cite{klumpp2005pnas,brenner2020sciadv}, $v_m = v_o( 1 - f/f_s)$, where $v_o$ is the load-free velocity of the motor. After each motor stepping event the force-balance condition is applied and new equilibrium values of $x,~y,~x_c$, and $y_c$ are obtained from Eqs. \ref{eq:xc}, \ref{eq:yc}, \ref{eq:surface} and \ref{eq:triangle}. This process continues until the motor unbinds from the MT, yielding estimates of the first passage time and detachment forces. We averaged all properties over $10^5$ independent simulation runs.

\begin{figure*}[t!]
    \centering
    \includegraphics[width=\linewidth]{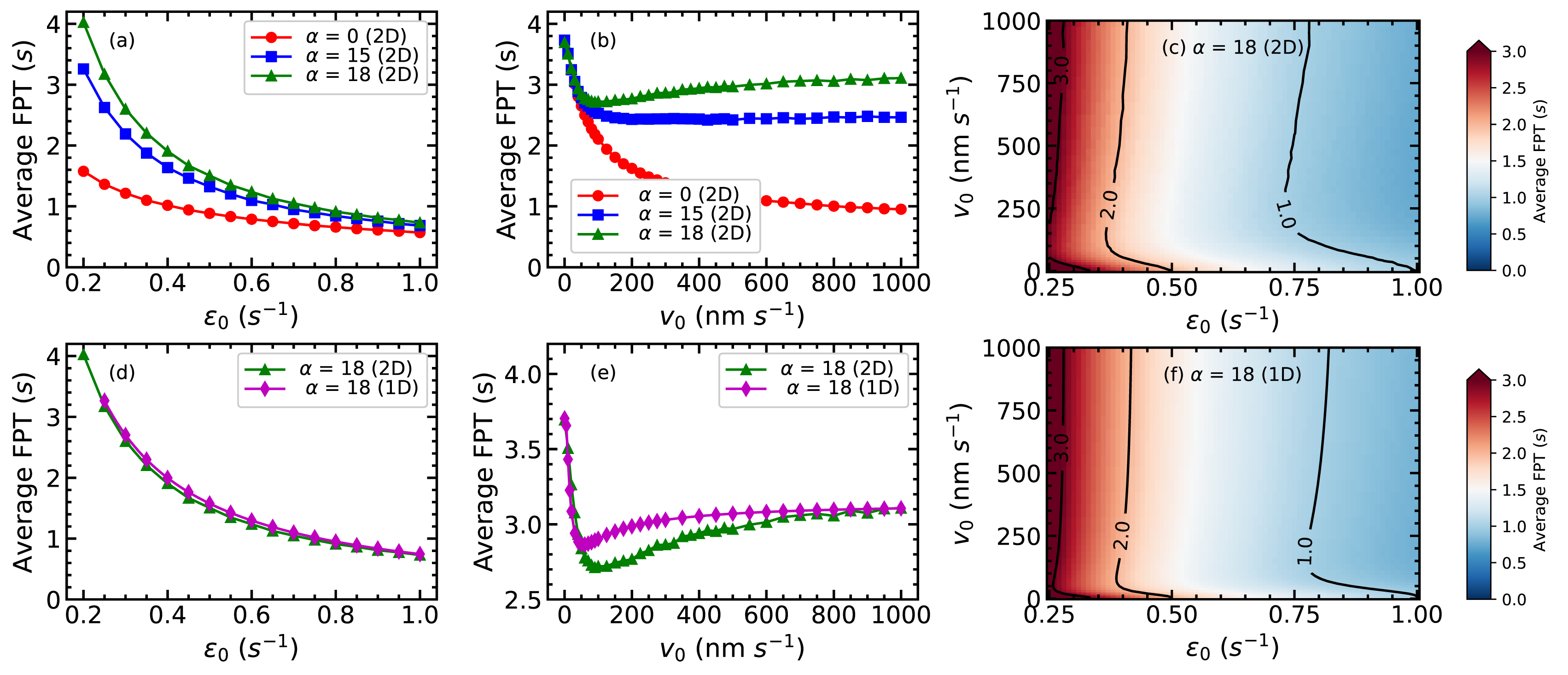}
    \caption{First passage time for single motor cargo transport using 2D approach. Panels (a) and (b) show the comparison between simulation-generated First Passage times $t$ for a single motor as a function of $\epsilon_o$ and $v_o$, respectively, for three different $\alpha$ values.  Panel (c) shows the corresponding contour map of variation of FPT with $\epsilon_o$  and $v_o$ simultaneously and demonstrates the appearance of a re-entrant under the influence of catch bond ($\alpha$ =  18  $k_BT$). The restlength of the motor, in this case, is chosen to be $l_o=55$ nm, and the radius of the cargo is assumed to be $R=500$ nm. Panels (d) and (e) show the comparison of results obtained from 2D approach vs 1D approach for $\alpha=18$. Panel (f) depicts the contour map of the average FPT in the case of 1D approach. In all of the plots for 1D approach restlength is chosen to be $l_o=0$. Rest of the parameters are as follows: $v_{o}$ = 362 nm s$^{-1}$ (for panels (a) and (d), $\epsilon_o$ = 0.27 s$^{-1}$ (for panels (b) and (e)), $f_{s}$ = 1.1 pN, $f_{d}$ = 0.67 pN, $f_{m}$ = 0.5  pN,  $f_{o}$ = 7 pN, $k_{t}=0.02$ pN nm$^{-1}$.}
    \label{fig:2D_phase}
\end{figure*}


For a cargo transported by a single motor in an optical trap, if the force at which catch bonding behaviour sets in for the motor, $f_m$ is the same or greater than the single motor stall force $f_s$, then both FPT and detachment force would be insensitive to the catch bonded nature of the dynein since the optical trap can only access force scale up to $f_s$ for a single motor. In principle, the allosteric deformation force scales which result in catch bonding behaviour in dynein need not be the same as the stall force of these motors. We thus explore the regime wherein,
$f_m < f_s$, in order to study the potential impact of catch bonding on cargo transport by a single motor. For a stall force of $f_s = 1.1 pN$ \cite{welte2004currbio,mallik2005currbio,hancock2014natrevmolcellbio}, we report results for a choice of $f_m = 0.5 pN$. We choose a trap stiffness of $k_t = 0.02 pN/nm $\cite{brenner2020sciadv}, which coupled with a dynein rest length of $l_o = 55 nm$ \cite{mckenney2010cell}, implies that the force in the vertical direction is such that the cargo transported by a single motor would never touch the MT in this scenario.



We characterise the First Passage Time (FPT) and investigate how different motor parameters such as the unbinding rate and motor velocities affect the FPT both in presence and in absence of catch bond. In an experimental context, motor velocity is a crucial parameter and can be controlled in various ways. Studies have shown that both kinesin and dynein motor velocity depends strongly on the ATP concentration \cite{schnitzer2000natcelbio,ross2006natcellbio,torisawa2014natcellbio,hirakawa2000pnas}. Interestingly, an in vitro motility assay study has also revealed that the velocity of single dynein-driven cargo transport varies exponentially with variation of temperature in accordance with Arrhenius equation \cite{hong2016biophysj}. Moreover, complexion of dynein with dynactin or BICD proteins and the size of the cargo transported by the dynein motor can affect the motor velocity significantly \cite{belyy2016natcellbio,schlager2014cellrep}. For example, the cytoplasmic dynein used in this study had a natural velocity of $\sim79~nms^{-1}$, while dynein-dynactin complex moves with a velocity $\sim513~nms^{-1}$. Conversely, the dynein velocity can rise upto $\sim257~nms^{-1}$ for a cargo of diameter $860~nm$ from $\sim50~nms^{-1}$ for a cargo of $20~nm$ diameter. Similarly, different dynein-activating proteins, such as dynactin, LIS1, NudE, Tau etc., can regulate the processivity of the dynein-driven cargo transport \cite{mckenney2010cell,schlager2014emboj,chaudhary2018traffic}. 

In Fig. \ref{fig:2D_phase} we show the variation of first passage time with unbinding rate and single motor velocity for a cargo being transported by a single motor. Fig. \ref{fig:2D_phase}(a) demonstrates that the first passage time for the bound-state of a cargo driven by a single motor decreases monotonically with an increasing principal unbinding rate $\epsilon_o$. Further, the blue and green curves demonstrate that in the case of catch bonded motors, the first passage time increases with catch bond strength for the given range of unbinding rates. Thus, a key tenet of catch bonded motors is improved persistence in the bound state of the cargo.
The standard response for slip-bonded motors transporting cargo against a harmonic force is a decrease in the persistence of the bound state as the cargo travels further away from the trap center. This is demonstrated in Fig. \ref{fig:2D_phase}(b) by the red curve where average first passage time decreases monotonically with the principal velocity parameter. This simply indicates that, with higher velocities, motors have a tendency to unbind faster as they venture into regions that elicit a stronger trap force. However, the blue and green curves demonstrate that catch bond counters this effect by decreasing the unbinding rate at these higher force regimes. There is an initial decrease in first passage time in the small $v_o$ limit as the cargo dynamics are not very likely to trigger a catch bond response at these velocities. However, at a sufficiently higher velocity, the probability of triggering catch bond grows and it is more typical for the cargo's bound state to have higher persistence, leading to an increase in First passage time with increasing $v_o$. We identify this re-entrant, exclusively, as a consequence of catch bond. We have also compared our results with the case of non-isotropic optical trap and both results match very well (see Suppl. Fig. 1). Note that while many of the experimental reports suggested that the usual residence time (First passage time in our case) of dynein is $\sim 1$s \cite{rai2013cell,brenner2020sciadv,leidel2012biophysj}, studies have shown that the residence time under optical trap for single dynein-driven cargo transport can be as large as $\sim5$s \cite{mallik2005currbio}, consistent with our theoretical results. Interestingly, complexing the dynein motor with LIS1 phosphoprotein can increase the residence time of single dynein significantly and can rise up to $\sim100$s \cite{mckenney2010cell}. The contour-lines in Fig \ref{fig:2D_phase}(c), further show that the re-entrant behavior of first passage time as a function of motor velocity is robustly exhibited for a wide range of principal unbinding rate $\epsilon_o$, in contrast with the slip-bond where it is completely absent (see Suppl. Fig. 2(a)).

\subsubsection{Force balance in one dimension}

We now turn to a simplified model of the cargo in this variable force optical trap setup, where we neglect motion along the vertical direction, and consider only the horizontal motion of cargo. The force balance condition then reads,

\begin{equation}
    k_t x_c = \Theta (x_m - x_c - l_o) k_m(x_m - x_c-l_o) 
     \label{eq:forcebalance}
\end{equation}
where $x_m$ is the displacement of the motor with respect to the center of the optical trap. It follows that, for a single motor, the force exerted on the motor is non-zero only when the motor distance from optical trap is greater than its rest length ($x_m > l_o$). The cargo position can be expressed as $ x_c =  \Theta (x_m - x_c - l_o) \left (\frac{k_m}{k_m + k_t}\right ) (x_m-l_o)$. Consequently, the cargo velocity, $v_c$, can be expressed in terms of the motor velocity, $v_m$ as, 
\begin{equation}
    v_c =  \Theta (x_m - x_c - l_o) \left(\frac{k_m}{k_m + k_t}\right) v_m
     \label{eqn:velocity}
\end{equation}
where the motor velocity depends linearly on the force, $v_m=v_o(1-f/f_s)$ \cite{klumpp2005pnas,brenner2020sciadv}. 
From. Eq.~\ref{eqn:velocity}, then, the instantaneous force $f$ exerted on the motor follows the evolution equation,
\begin{equation}
\frac{df}{dt}=k_t v_c=\Theta (x_m - x_c - l_o) \left(\frac{k_t k_m}{k_m + k_t}\right) v_o \left(1-\frac{f}{f_s}\right) \label{eq:f_dot}
\end{equation}
We can thus relate the instantaneous force $f$ to the instantaneous cargo position $x_c(t)$ by integrating Eq. \ref{eq:f_dot} and it can be expressed in terms of the time interval $t$ for which the motor is bound to the MT filament as, 
\begin{equation}
        f(t) =k_t x_c(t) = \Theta (v_ot-l_o) f_s \left[ 1 - \exp\left( -\frac{k_t k_m (v_ot-l_o)}{(k_t + k_m)f_s} \right)\right ] 
     \label{eq:force}
\end{equation}
where the $\Theta$ function is non-zero only for $t$ greater than the mean time taken by the motor to walk beyond its rest length.

For non-catch bonded motors, we have $ \epsilon = \epsilon_{o} e^{ k_t x_c/f_k}$. Using Eq.(\ref{eq:force}), we obtain,

\begin{equation}
    \epsilon (t) = \epsilon_{o} \exp \left[ \Theta (v_o t - l_o) \frac{f_s}{f_k} \left( 1 - e ^{-K(v_ot-l_o)}\right )\right ]
    \label{eq:ku}
\end{equation}
where $K = \frac{k_t k_m}{(k_t + k_m)f_s}$. For the catch bonded motors, the form of the unbinding rate is obtained from Eqs.(\ref{eq:catch-ku}) and (\ref{eq:deform}), with the instantaneous force being given by Eq.~\ref{eq:force}. The unbinding rate is then substituted in the integral form of Eq.(\ref{eq:S(t)}), to obtain the survival probability distribution and hence the probability distribution of the first passage time. Once we obtain the first passage time distribution, we can calculate the detachment force distribution, $P(f)$, using Eq.~\ref{eq:P_f}.

In addition we also stochastically simulate cargo trajectories in this 1D configuration (see Suppl. Sec. 2 for details). We simulate the trajectories and obtain the statistical measures for multiple motors by averaging over an ensemble of 5000 runs of the process. 

We first compare the variation of the first passage time as a function of the principal (load-free) unbinding rate obtained from this one-dimensional analysis with the results of the full two-dimensional treatment. As can be seen in Fig.~\ref{fig:2D_phase}(d), the two curves effectively overlap, showing that the 1D treatment can effectively capture the catch bonded transport ($\alpha=18k_BT$). In Fig.~\ref{fig:2D_phase}(e), we show the variation of the FPT with the load-free velocity of the motor. Even in this case, the 1D treatment is able to accurately capture the re-entrant behaviour with increasing $v_0$ as was seen in the 2D case with an excellent quantitative match between the two curves (see Suppl. Fig. 3 for $\alpha=0~k_BT$ and $\alpha=15~k_BT$ cases). This is a generic feature, as can be seen in Fig.~\ref{fig:2D_phase}(f), where the contour plots show the re-entrant variation of FPT in the $v_0$-$\varepsilon_0$ plane. For the sake of completeness, we also simulated single motor transport in 1D and found that the simulation results match excellently with the analytical results (see Suppl. Fig. 4). Having demonstrated that the 1D treatment can capture the behaviour of FPT, for simplicity, we now continue with our 1D approach for the rest of the manuscript.

\begin{figure*}[t]
    \centering
    \includegraphics[width=\linewidth]{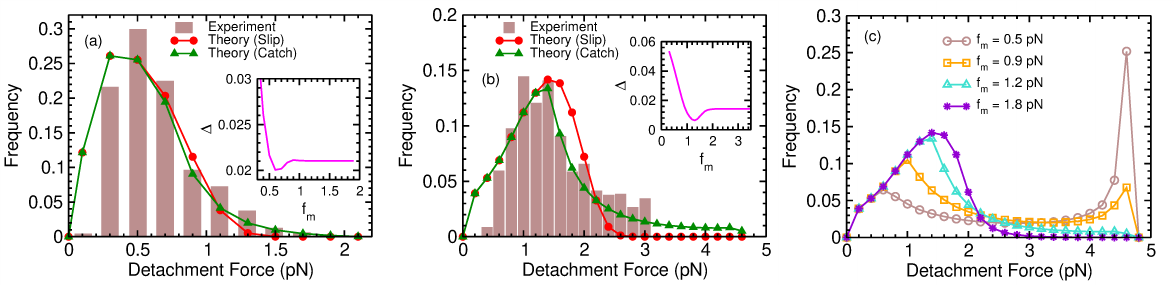}
    \caption{Comparison of our 1D model with experimental data. Panel (a) and (b) compares detachment force distribution between theory and experiment \cite{brenner2020sciadv} for trap stiffness $k_t=0.01$ pN/nm and $k_t=0.1$ pN/nm respectively. Note that the experimental histogram in panel (b) is plotted till 3pN due to unavailability of detachment force data\cite{brenner2020sciadv}. Panel (c) shows the variation of the probability distribution of detachment force for different $f_m$ for catch bonded motors ($\alpha=18)$. The values of $f_m$ in panels (a) and (b) are obtained from the least square fitting of theoretical curves with experimental data. The variation of the sum of squared residuals ($\Delta$) is shown as a function of $f_m$ in the insets of panels (a) and (b). The optimum values of $f_m$ in two cases are (a) $f_{m}$ = 0.6  pN and (b) $f_{m}$ = 1.3  pN. In panel (a) the stall forces are chosen to be $f_s=2$ pN while in panels (b) and (c) stall force is chosen to be $f_s=4.5$ pN. Rest of the parameter values used are as follows : $\epsilon_o$ = 6 s$^{-1}$ \cite{brenner2020sciadv}, $v_{o}$ = 540 nm s$^{-1}$ \cite{brenner2020sciadv},  $l_{o}$ = 0 nm, $f_{d}$ = 0.67 pN,  $f_{o}$ = 7 pN, $\alpha=18$ (for catch bond). }
    \label{fig:experimental_fit}
\end{figure*}

\subsection{Estimate of force scale for onset of catch bond}
For typical experimental settings, stalling event of the cargo occurs for a range of forces. Associated with the distribution of the stall forces, one can define an {\it average} stall force, which is usually invoked as an experimentally reported measure of $f_s$. The full distributions are often very broad, and the {\it maximal} observed stall force can be more than twice the average stall \cite{brenner2020sciadv}. Recent optical trap experiments on human dynein purified from human embryonic kidney (HEK) 293 cells have shown that for this weakly processive mammalian dynein, for weak trap stiffness, typically the motor unbinds much before it experiences the {\it maximal} stall force. \cite{brenner2020sciadv}. This in turn implies the experimentally reported measure of average stall force is a function of trap stiffness. 
We compare and contrast the predictions of the theoretical model with the experimental data in the presence and absence of catch bond, and the dependence of our theoretical results on the force scale $f_m$ where the catch bond sets in. The optimal force scale $f_m$ is chosen as that value which minimises the sum of least square differences ($\Delta$) between the experimental data and the theoretical predictions.
For the current case of a cargo transported by a single motor, we first characterise the probability distribution of detachment forces obtained by our theoretical approach with the experimental data for two  different values of the trap stiffness - $k_t=0.01 pN$ (Fig. \ref{fig:experimental_fit}(a)), and $k_t=0.1 pN$ (Fig. \ref{fig:experimental_fit}(b)) - for catch bonded and non-catch bonded motors ( Eq. \ref{eq:P_f}). 
While the experiments observe the full distribution of stall forces, spanning an entire range of detachment force scales, for the purpose of comparison of our theoretical model with experiments, we assign the maximum value of stall force that is observed in the experiment as the choice of fixed $f_s$ that we use in our theoretical model.


For a relatively low value of trap stiffness $k_t  = 0.01 pN/nm$ (Fig. \ref{fig:experimental_fit}(a)), corresponding to an experimentally observed maximal stall force of $2 pN$, the peak of theoretical probability distribution is largely insensitive to the catch bonded nature of dynein bond. However, the tail of the distribution is longer for the catch bonded case in comparison to the non-catch bonded case. Indeed, the tails of the experimentally observed probability distribution match better with the theoretically obtained probability distribution of detachment forces for the catch bonded case with a catch bond threshold force scale $f_m = 0.6 pN$. Note that the average stall  force at this trap stiffness is $f_s = 0.9 pN$ \cite{brenner2020sciadv}. It's worthwhile to point out that when the trap stiffness is low, except for the tail of the distribution function, the probability distribution of detachment force for the case of catch bonded motors is virtually indistinct from the probability distribution for non-catch bonded motors. This may be understood in the following manner. When the trap stiffness is low, most of the unbinding events happen typically before the cargo attains the displacement which is larger than $f_m/k_t$ which results in the theoretical probability distribution being virtually indistinguishable from the non-catch bonded case. The theoretical curves at this trap stiffness - both for the catch bonded and non-catch bonded cases - corresponding to the probability distribution exhibit broad qualitative agreement with the experimentally obtained distribution for this case. Note that in the experiments, only displacements greater than a certain threshold were recorded, implying that data was not available for very low detachment forces. This leads to a discrepancy between the observed and theoretical frequency distributions at very low detachment forces as seen in Fig. \ref{fig:experimental_fit}(a) and (b). 
For high values of the trap stiffness, ($k_t = 0.1 pN/nm$), the theoretically obtained distribution for the non-catch bonded case exhibits significant divergence from the experimental curve, as shown in Fig.~\ref{fig:experimental_fit}(b). In contrast, the theoretically obtained distribution which incorporates the catch bond mechanism exhibits a very good match with the experimental curve for $f_m=1.2 pN$. The average stall force in this case is $f_s = 1.9pN$. Thus, our analysis suggests that the catch bond  sets in at a force scale that is lower than the average stall force.

The sums of squared residuals ($\Delta$) are shown in the insets of Fig. \ref{fig:experimental_fit}(a) and (b) for different values of $f_m$. In principle, the fit to the data can be tuned by the force scale $f_m$ at which the catch bond sets in, as shown for four $f_m$ values in Fig.~\ref{fig:experimental_fit}(c). For very low values of $f_m$, the catch bond sets in much earlier and leads to a pronounced unphysical peak in the distribution. On the other hand, for relatively large values of $f_m$, there is no distinction between the catch and slip bond cases. Thus, the physically allowed ranges of $f_m$ which correspond to the experimentally observed distributions constrain the force scale at which catch bond sets in.

\begin{figure*}[t]
    \centering
    \includegraphics[width=\linewidth]{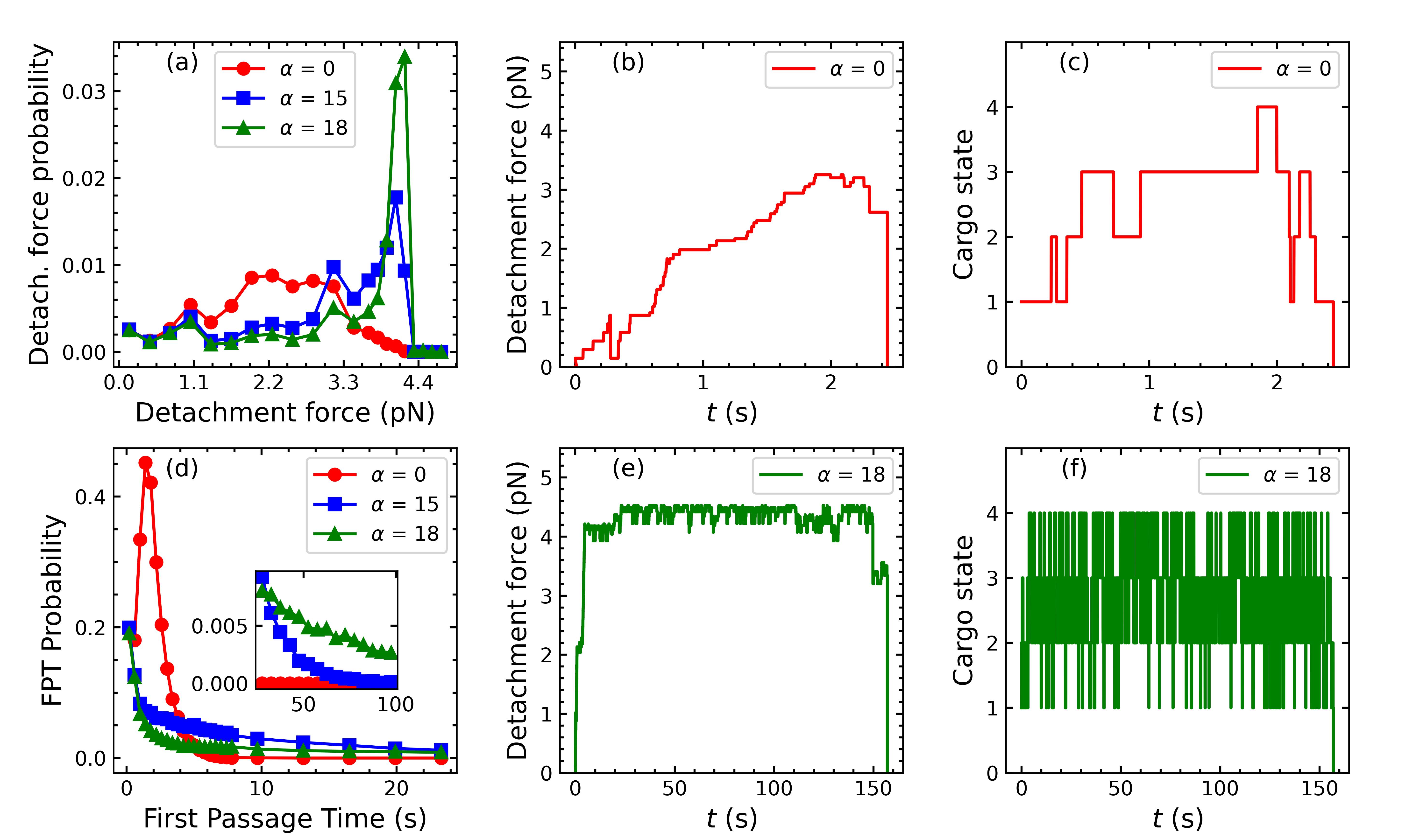}
    \caption{Effects of catch bond on transport properties of a team of dynein motors. Panels (a) and (d) show the normalized probability distributions for Detachment force $f$ and First Passage time $t$ respectively of the cargo for $N = 4$. Red curves correspond to slip bond ($\alpha$ = 0 $k_{B}T$) , blue curves to catch bond ($\alpha$ = 15 $k_{B}T$) and green curves to Strong catch bond ($\alpha$ = 18 $k_{B}T$). The inset in panel (d) shows a magnified view of the distribution corresponding to longer time scales. Panels (b) and (e)  show sample trajectories of the cargo in the absence and presence of catch bond respectively. Panels (c) and (f) show the corresponding trajectories of the cargo in state-space in the absence and presence of catch bond respectively. The parameter values used are as follows : $\pi_o$ = 1.6 s$^{-1}$, $\epsilon_o$ = 0.27 s$^{-1}$, $l_{o}$ = 55 nm, $v_{o}$ = 362 nm s$^{-1}$, $f_{s}$ = 1.1 pN, $f_{d}$ = 0.67 pN, $f_{m}$ = 1.1 pN,  $f_{o}$ = 7 pN, $k_{t}=0.02$ pN nm$^{-1}$.}
    \label{fig:traj}
\end{figure*}

\subsection{Cargo transport by multiple motors}

Having established the significance of catch bonded response for single motors, we now turn to the consequences of catch bond for multiple motors. For single motors, $f_m$ must necessarily be less than the maximal stall force to observe catch bond effects. In the previous section, we have shown that the estimated value of $f_m$ is lower than the average stall force $f_s$. However, this difference between these two force scales is not as critical for the case of transport by multiple motors since catch bond effects manifest themselves even in the limiting case of $f_m = f_s$. Thus for the sake of simplicity of analysis for transport of cargo by multiple motors, we set $f_m = f_s$. Note that all catch bond mediated properties will be even further magnified on choosing $f_m < f_s$, and hence this serves as a minimal prediction for catch bonded transport. For studies of motor-driven transport, usual optical trap stiffness are in the regime $0.01 -0.03$ pN/nm \cite{rai2013cell,brenner2020sciadv}. Throughout the rest of the manuscript, we restrict ourselves to a physical realistic trap stiffness of 0.02 pN/nm. At this value of trap stiffness, the mean stall force is $f_s = 1.1 pN$ \cite{brenner2020sciadv} and hence we choose $f_m = 1.1 pN$. A catch bond response can still be triggered in this regime, for multiple motors, as the cargo can be transported beyond the distance corresponding to the stall force of a single motor. 

\begin{figure*}[t]
    \centering
    \includegraphics[width=\textwidth]{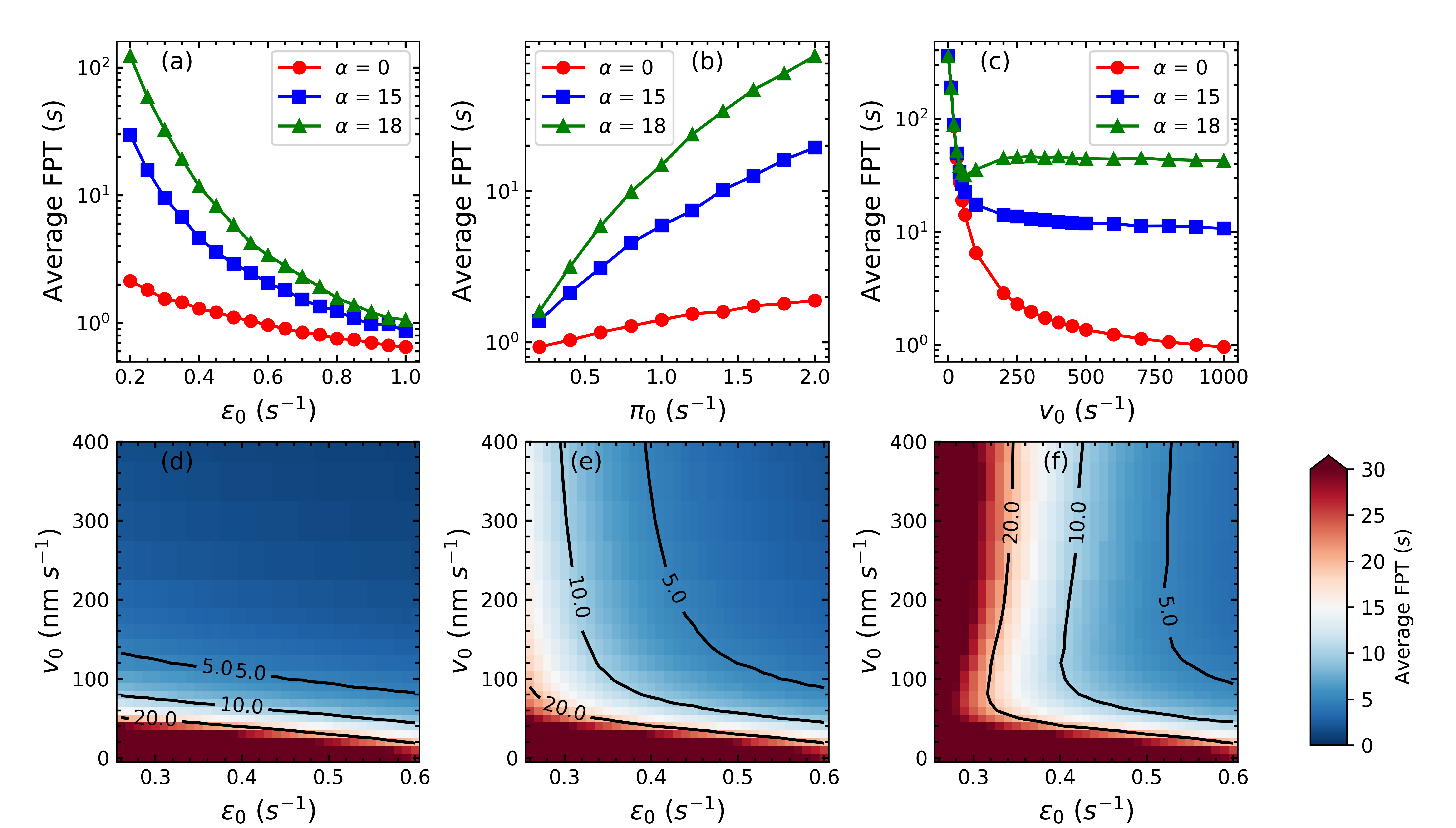}
    \caption{Variation of average FPT as a function of various motor properties. Panel (a) denotes the average First Passage time (FPT) as a function of $\epsilon$, panel (b) denotes the average First Passage time as a function of $\pi$, and panel (c) shows average First Passage time as a function of $v_{o}$. Red corresponds to slip bond ( $\alpha = 0 k_{B}T$), blue to catch bond ($\alpha$ = 15 $k_{B}T$) and green to strong catch bond ($\alpha$ = 18 $k_{B}T$). Panels (d),(e) and (f)  are contour plots of Average First Passage Time in the $\epsilon$-$v_{0}$ plane, with slip bond ($\alpha$ = 0 $k_{B}T$), presence of catch bond ($\alpha$ = 15 $k_{B}T$) and strong catch bond ($\alpha$ = 18 $k_{B}T$) respectively. The parameters used are: N = 4, $\pi_o$ = 1.6 s$^{-1}$ (for panel (a)), $\epsilon_o$ = 0.27 s$^{-1}$ (for panel (b)), $v_{o}$ = 362 nm s$^{-1}$. $l_{o}$ = 55 nm. $f_{s}$ = 1.1 pN, $f_{d}$ = 0.67 pN, $f_{m}$ = 1.1 pN, $f_{o}$ = 7 pN, $k_{m}$ = 0.2 pN nm$^{-1}$, $k_{t}$ = 0.02 pN nm$^{-1}$. }
    \label{fig:pi-e-v0}
\end{figure*}

\subsubsection{Catch bond alters transport properties}
In Fig.\ref{fig:traj} we see a comparison between trajectories of a cargo carried by a maximum of 4 motors based on the presence or absence of catch bond. We observe in  Fig. \ref{fig:traj}(a) that both detachment force distributions terminate at $\sim 4.4 pN$ which corresponds to the detachment force $f_{max} = Nf_{s}$, the maximum possible force that the team of motors can generate 
before all N motors are stalled. Although in case of slip bond ( red curve ) a significant fraction of runs result in detachment at intermediate loads 
(in between $1.5-3.5pN$), the detachment force distribution for catch bond (blue and greens curves) is biased towards larger distances ($\geq$ 4 $pN$). We also observe characteristic peaks at intervals of $\sim 1.1 pN$ for both slip and catch bond (Fig.\ref{fig:traj}(a)). These peaks correspond to the external force that equals integer multiples of the stall force of a single motor, and indicate that most of the unbinding events take place at superstalled state of the motors. 
In both cases, this can be understood by the following argument: due to relatively high motor velocity ($v_0$), the motors most often reach a load force equivalent to multiples of stall force (number of bound motors times individual motors’s stall force) before they undergo any unbinding event. Once the load force on each bound motor is equal to their individual stall force, the cargo halts until another motor binds to the MT and starts sharing the load or until all motors detach from that position. This high residence time at load forces equal to multiples of $f_s$ increases the probability of cargo unbinding events at those load forces which results in peaks. The difference between slip and catch bond is that in slip bond the peaks at higher multiples of stall force diminishes as the motors do not often stay bound long enough for the cargo to reach that load force regime. Catch bond counters this effect with reduction of unbinding rate at superstalled load which increases the probability of the cargo remaining bound until the next motor binding event and enhances the probability of the cargo to reach a higher load force regime. Therefore in the presence of catch bond the peaks at higher multiples of stall force are also enhanced.

A similar persistent nature for catch bond can be spotted from the comparison of First Passage time (FPT) distribution curves for two cases (Fig.\ref{fig:traj}(d)). While most of the runs terminate after smaller duration (about a few seconds) for slip bond (red curve), the catch bond-associated distributions (blue and green curves) have very long tails extending to larger timescales (a few minutes). The tail is longer for stronger catch bonded motors (higher $\alpha$) as shown in Fig.\ref{fig:traj}(d) (inset). Fig.\ref{fig:traj}(b) and Fig.\ref{fig:traj}(c) show a representative trajectory of the cargo for slip bonded motors in terms of instantaneous force on the cargo and number of motors attached to the filament (motor state of the cargo) as a function of time respectively while Fig.\ref{fig:traj}(e) and Fig.\ref{fig:traj}(f) depict the same for catch bonded motors. In higher motor states, the cargo can travel farther away from the trap-center and hence be subjected to a larger load force. As depicted in Fig. \ref{fig:traj}(c), in the slip bond case, this results in a rapid unbinding of motors as with each unbinding, the remaining motors take on a bigger share of load, exponentially increasing their tendency to unbind. However, in the catch bond case, beyond the threshold force ($f_m$) the unbinding rate decreases with increasing load. Thus catch bond helps the motors to stay bound in a higher load regime until unbound motors (if any) rebind to the filament again. This results in frequent occurrence of cargo transitions from a lower to a higher motor state over the course of the cargo transport as depicted in Fig. \ref{fig:traj}(f).\\

\subsubsection{Motor Velocities and Binding/ Unbinding rates strongly affect cargo transport characteristics}


 Similar to the case of a single motor, for multiple motors, the average FPT monotonically decays with increasing unbinding rate (Fig. \ref{fig:pi-e-v0}(a)) while it monotonically increases with increasing binding rate (Fig. \ref{fig:pi-e-v0}(b)). We also observe that the presence of catch bond (blue and green curves) causes the average first passage time to differ by an order of magnitude from that of slip bond (red curve) for biologically relevant binding and unbinding rates. 
 
\begin{figure*}[t]
    \centering
    \includegraphics[width=0.85\linewidth]{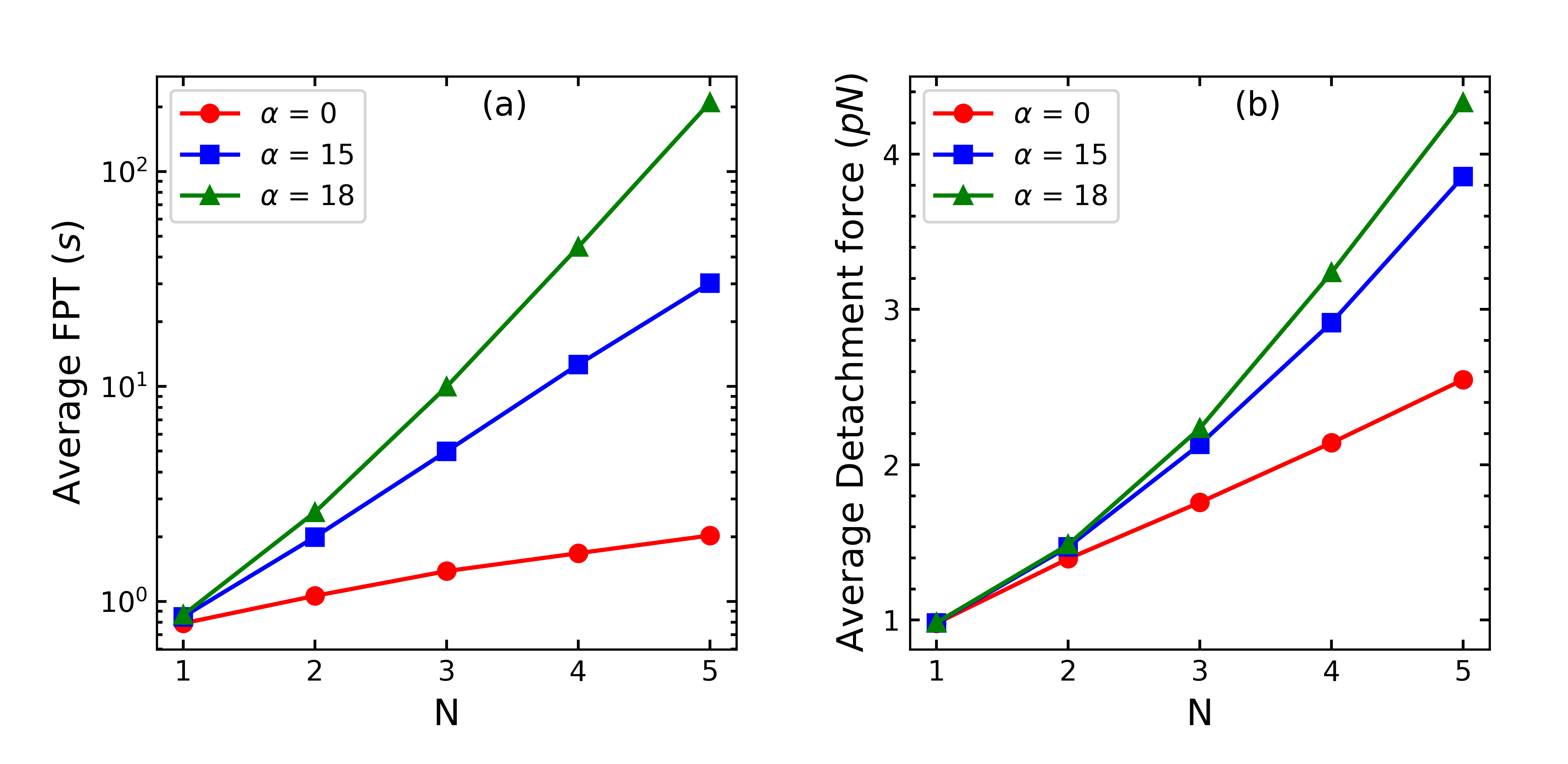}
    \caption{Effects of varying number of motors on transport properties. Panel (a) depicts the average FPT as a function of total number of motors N and panel (b) shows the average detachment force with varying N.  Red corresponds to slip bond ($\alpha = 0 k_{B}T$), blue corresponds to catch bond  ($\alpha = 15 k_{B}T$) and green with strong catch bond ($\alpha = 18 k_{B}T$). The parameters used are : $\pi_o$ = 1.6 s$^{-1}$, $\epsilon_o$ = 0.27 s$^{-1}$ , $v_{o}$ = 362 nm s$^{-1}$. $l_{o}$ = 55 nm. $f_{s}$ = 1.1 pN, $f_{d}$ = 0.67 pN, $f_{m}$ = 1.1 pN, $f_{o}$ = 7 pN. $k_{m}$ = 0.2 pN nm$^{-1}$ and  $k_{t}$ = 0.02 pN nm$^{-1}$. }
    \label{fig:N-v0}
\end{figure*}

As previously  noted for single motor-driven cargo, higher velocities imply that the motors reach a higher load regime faster and thus become more prone to unbinding in case of cargo transported by a team of dyneins (Fig.\ref{fig:pi-e-v0}(c)). Therefore the average FPT associated with faster motors is lower than that of the slower motors.  In Fig.\ref{fig:pi-e-v0}(c), we observe a re-entrant behaviour in the average first passage time with increase in $v_{o}$ for strong catch bond ($\alpha=18$) (green curve). The average first passage time initially decreases with increasing $v_{o}$ until it reaches a minimum at $v_o\sim$70 $nm~s^{-1}$ and then again starts to increase until $v_o\sim$300 $nm~s^{-1}$. Beyond $v_o\sim300 nms^{-1}$ the average FPT decreases very slowly, in contrast to the single motor behaviour in Fig.~\ref{fig:2D_phase}(e) where the FPT saturates with velocity for the case of strong catch bond. The maximum can be ascribed to the threshold where the fraction of cargo trajectories undergoing catch bonded states is maximized. For even higher velocities the effect of catch bond remains the same. However, the lagging motors, in this case, catch up to leading motors quickly and hence reduce the higher load on the leading motors. This reduces the characteristic time-scale for the superstall state of the leading motors. This leads to a very small decrease in the first passage time for very high motor velocity. Fig. \ref{fig:pi-e-v0}(d-f) depicts contour maps of average first passage time in the $\epsilon_o - v_{o}$ plane under slip bond and two different  catch bond strength ($\alpha$ = 15 $k_BT$ and $\alpha$ = 18 $k_BT$ respectively). We observe that similar to Fig.\ref{fig:pi-e-v0}(c) (green curve) there is a re-entrant behaviour between $\epsilon_o$ = 0.25 and 0.5 as we increase $v_o$ (Fig.\ref{fig:pi-e-v0}(f)). \\

\subsubsection{Effect of changing number of motors}
Fig.\ref{fig:N-v0} shows that the average first passage time and average detachment force increase monotonically with the total number of motors (N). This is due to the fact that higher number of motors results in a lower share of load force for individual motors and hence the unbinding rates of the individual motors remain lower than in the case of lower values of N. While both the average FPT and average detachment force in slip bond and catch bond are comparable for low motor numbers, for N=7, catch bond values differ significantly from slip bond values for higher N reaching almost two orders higher in the case of average FPT and twice the value in the case of average detachment force. It is noteworthy that the catch bond strengthens drastically as N increases. This is due to the fact that a higher number of motors implies that the cargo can travel further away from the trap center before all motors are stalled, generating large external force. As motors progressively unbind, this high external load is shared among the remaining motors leading to a large individual super-stall load and stronger catch bond.  This is substantiated by the increase in the average first passage time by orders of magnitude with an increase in N due to catch bond. \\   


\begin{figure*}[t]
    \centering
    \includegraphics[width=\linewidth]{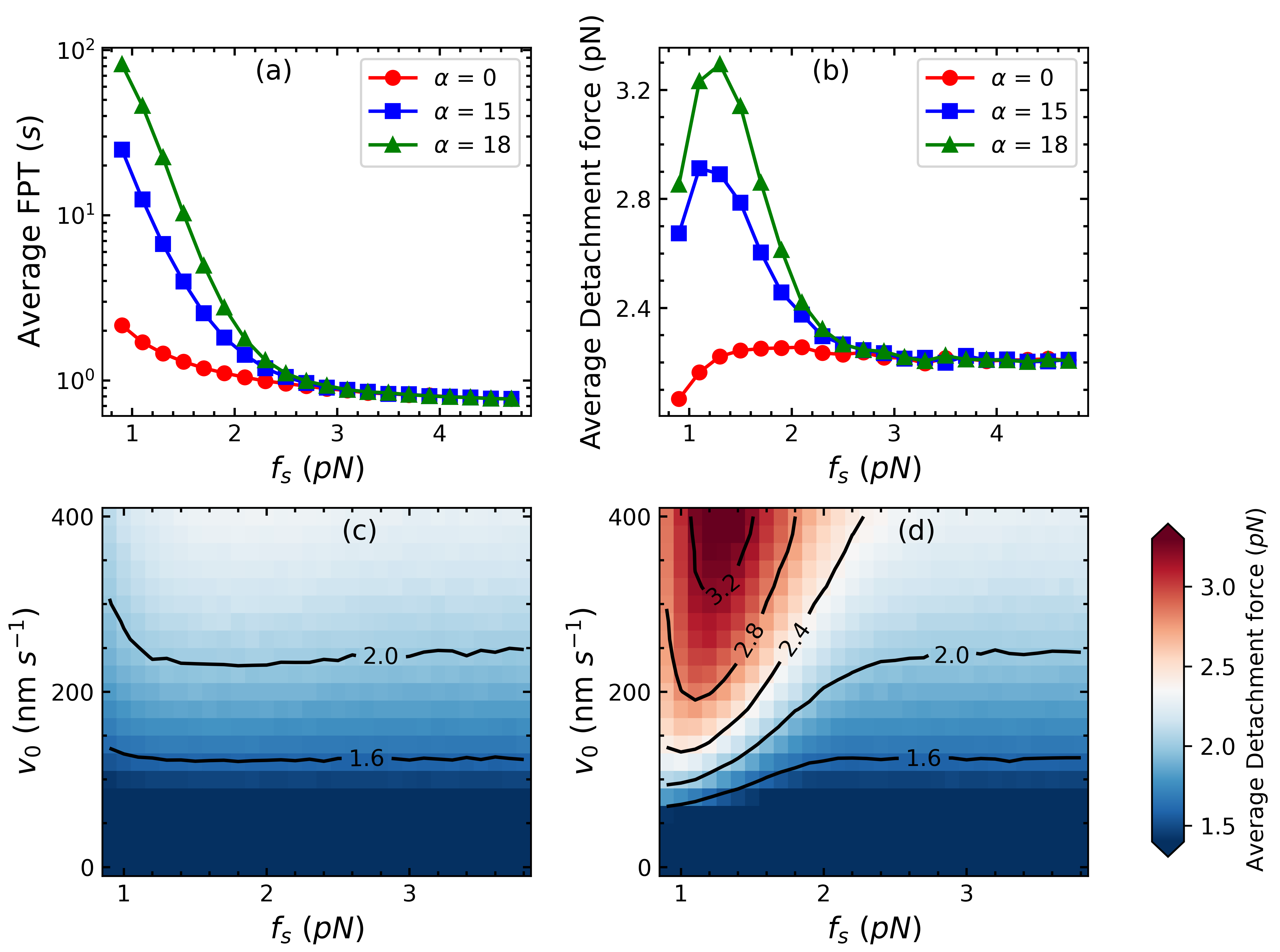}
    \caption{Effects of varying stall force on average FPT and average detachment force. Panels (a) and (b) shows the variation in average FPT and detachment force with stall force $f_{s}$. Panels (c) and (d) represent contour plots of average detachment force in the $f_{s}-v_{o}$ plane in the absence  ($\alpha = 0~ k_{B}T$ ) and presence of catch bond ($\alpha = 18 ~k_{B}T$ )  respectively.  For panels (a) and (b), red, blue, and green correspond to slip bond, catch bond, and Strong catch bond respectively. The parameters used are: N = 4, $\pi_o$ = 1.6 s$^{-1}$ , $\epsilon_o$ = 0.27 s$^{-1}$ , $v_{o}$ = 362 nm s$^{-1}$. $l_{o}$ = 55 nm. $f_{d}$ = 0.67 pN, $f_{m}$ = $f_{s}$, $f_{o}$ = 7 pN. $k_{m}$ = 0.2 pN nm$^{-1}$, $k_{t}$ = 0.02 pN nm$^{-1}$. }
    \label{fig:fs-v0}
\end{figure*}

\subsubsection{ Effect of variation of stall force}
The stall force of dynein is a well-debated topic. While most experiments suggest that cytoplasmic dynein has a stall force of 1.1 pN \cite{welte2004currbio,mallik2005currbio,hancock2014natrevmolcellbio}, it has been reported that the stall force can be modified to $\sim$ 4 pN by complexing the dynein with dynactin and BicD2 (DDB complex) \cite{belyy2016natcellbio}. On the other hand, dyneins found in mammals and yeasts are reported to have a very high stall force ($\sim$ 7 pN) \cite{toba2006pnas,gennerich2007cell}. Experiments have also reported variations in stall forces due to changes in the ATP concentration \cite{singh2005monte}. Further, varying the trap stiffness in optical experiments can drastically alter the observed mean stall forces \cite{brenner2020sciadv}. Fig.\ref{fig:fs-v0}(a) and Fig.\ref{fig:fs-v0}(b) show the variation of average first passage time and average detachment force respectively with increasing stall force ($f_{s}$). The average detachment force increases initially with $f_{s}$, reaches a maximum, and then decreases with stall force for both slip and catch bonds. The rise is seen to be steeper in the catch bond case (blue and green curves). As the average detachment force decreases, both slip bond and catch bond appear to converge gradually at higher $f_{s}$ ($\geq 2.5$ pN). This indicates that since $f_{m}$ increases with $f_{s}$, the likelihood of a motor achieving a catch bond state drops with increasing $f_s$. This is also reflected in the average first passage time. Average FPT decreases rapidly with increasing $f_{s}$ for both slip bond and catch bond and the curves converge at higher $f_{s}$. The decrease in average FPT with increasing $f_{s}$ is a consequence of higher velocities arising because $v = v_0 (1- f/f_s)$ increases as $f_s$ increases, leading to further displacement from the trap centre and hence higher unbinding. In the case of detachment forces, for intermediate values of $f_s$, motors can travel further and do not unbind as rapidly resulting in a marginally higher run length (and hence detachment force). Catch bond exaggerates this effect by promoting further displacement by increasing the persistence of the cargo's bound state at higher loads. Catch bond, however, is triggered with a smaller likelihood with the increase of $f_s$ and therefore, ceases to be of any aid beyond a certain force scale. This effect is consistently reproduced over a range of higher $v_0$ ( 100 - 400 $nm  s^{-1}$) (Fig.~\ref{fig:fs-v0}(c) and \ref{fig:fs-v0}(d) for the slip and catch bond cases). Detachment forces can vary by almost $1 pN$ in the presence of catch bond over the range of stall forces investigated, while there is no appreciable variation for the case of slip bond.

\section {Discussion}


One of the characteristic features exhibited by the dynein motors is known as catch bond. Grasping the implications of this property of the dynein motors for the transport properties of motor-driven transport of cellular cargo and mechanical property of motor-cellular filament complexes is yet to be fully appreciated.

An important parameter for catch bonds is the force scale $f_m$ at which the catch bonding effect starts to set in. Our theoretical analysis provides a plausible estimate of this force scale, obtained through a careful reexamination of experimental data on single dynein-driven transport \cite{brenner2020sciadv}. Since a complete molecular understanding of the structural mechanism of the dynein catch bond is still absent, it is not possible to estimate this force scale empirically. Experimental data on unbinding rates suggest that this force scale is comparable to the stall force of the motor $f_s$, and previous studies on cooperative transport have investigated the regime where these forces are equal $f_s = f_m$ \cite{kunwar2011pnas,rai2013cell,nair2016pre,puri2019prr}. However, this becomes important in particular, for the case of transport of cellular cargo by a single motor, where the difference between catch bonded and non-catch bonded case manifests itself {\it only if}, catch bonding in the motor sets in at a force scale which is lower than the characteristic detachment force of the motor. The interpretation in this case is complicated by experimental evidence that the  observed average stall force can depend on the trap stiffness and is in general much lower than the maximal stall force. Using careful comparisons with experimental data for detachment forces for cargo transported by single dynein \cite{brenner2020sciadv}, our work suggests that indeed, the catch bond sets in at a force scale $f_m$ which is lower than the average stall forces at a given trap stiffness, $f_m \leq f_s$. The signature of catch bonding manifests as increased probabilities for higher detachment forces (Fig.~\ref{fig:experimental_fit}b). Given that these force scales are distinct, possible experiments that can differentially impact these two forces scales, possibly through complexation with dynein adaptors \cite{hancock2014natrevmolcellbio}, can then show drastic deviations of detachment force distributions from the  non-catch bonded case.

For single motor driven transport, we analyse motion in an optical trap within both a two-dimensional framework which accounts for the vertical restoring force from the optical trap, and a one-dimensional framework, as is common in the literature \cite{kunwar2011pnas,bhat2016epje,nair2016pre,puri2019prr,ohashi2019traffic}. We show that for biologically relevant parameter regimes, the one-dimensional description captures the essential physical properties of the system. 



Experiments performed with motor teams carrying single phagosomes have demonstrated that while team of kinesin motors fails to work collectively under conditions of variable force setting of optical trap, team of dynein motors is not only able to counteract larger forces ( due to the optical trap) but their tenacity is much more pronounced \cite{rai2013cell,mallik2013ticb}. Indeed the findings of our model are consistent with these experimental observations where catch bonded transport manifests in increased lifetimes of cargo. Interestingly, we observe that as a function of the motor velocity, the average first passage time shows a re-entrant behaviour even for the case of cargo carried by single dyneins. Such counterintuitive phenomena become even more robust when we consider collective effects due to transport by multiple motors. In this case, we also observe a `re-entrant' like behaviour where we find that average FPT of transported cargo initially decreases before again increasing with the increase in motor velocity. We also observe a non-trivial dependence of the average detachment force with the dynein motor stall force $f_s$. In particular, we find that for certain parameter regimes, on increasing the stall force $f_s$, the average detachment force initially increases before again decreasing at higher values of stall force. Both motor velocities and stall force are known to be modulated by cells through a variety of mechanisms, particularly through complexation with dynein adaptors \cite{ohashi2019traffic,belyy2016natcellbio,ross2006natcellbio,torisawa2014natcellbio,hirakawa2000pnas,hong2016biophysj,schlager2014cellrep,schlager2014emboj,chaudhary2018traffic}. Our work thus shows that for catch bonded motor-driven transport, molecular adaptations can help the cell generate complex transport behaviour which can have functional implications for intracellular transport.

\section*{Author Contributions}
NS carried out simulations, analyzed model and data and wrote manuscript, SG formulated and carried out theoretical calculations, analyzed model and data and wrote manuscript,  MKM and SM designed study, analyzed model and data and wrote manuscript.

\section*{Conflicts of interest}
Authors declare no conflict of interests.

\section*{Acknowledgments}
Financial support is acknowledged by SM, NS and MKM for SERB project No. EMR /2017/001335. MKM and SG acknowledge financial support from IIT Bombay.

\bibliography{refs} 
\bibliographystyle{unsrt} 


\end{document}